\documentclass[%
 reprint,
 amsmath,amssymb,
 aps,
]{revtex4-2}

\usepackage{graphicx}% Include figure files
\usepackage{dcolumn}% Align table columns on decimal point
\usepackage{bm}% bold math
\begin{document}

\title{Magnetoconductance evolution across the topological--trivial phase transition in ${In_{x}}({Bi_{0.3}}{Sb_{0.7}})_{2-x}{Te_3}$ thin films}

\author{Sambhu G Nath}
\altaffiliation{Department of Physical Sciences, IISER Kolkata}
\author{Subhadip Manna}
\altaffiliation{Department of Physical Sciences, IISER Kolkata}
\author{Kanav Sharma}
\altaffiliation{Department of Physical Sciences, IISER Kolkata}
\author{Amar Verma}
\altaffiliation{School of Physical Sciences, UM-DAE Centre
for Excellence in Basic Sciences, University of Mumbai,
Mumbai 400098, India}
\author{Ritam Banerjee}
\author{R K Gopal}
\author{Chiranjib Mitra}
\email{Corresponding author:chiranjib@iiserkol.ac.in}
\affiliation{Indian Institute of Science Education and Research Kolkata,Mohanpur 741246, West Bengal, India}
\date{\today}

\begin{abstract}
We investigate the evolution of electronic transport across the topological-trivial phase transition in ${\rm In}_{x}({\rm Bi}_{0.3}{\rm Sb}_{0.7})_{2-x}{\rm Te}_3$ thin films by systematically tuning the indium concentration $x$. Increasing $x$ reduces the effective spin-orbit coupling, driving a topological quantum phase transition near $x \approx 7\%$, and at higher disorder a crossover from diffusive to strongly localized transport around $x \approx 15\%$. In the diffusive regime, the magnetoconductance is well described by the Hikami--Larkin--Nagaoka formalism, with the evolution of the WAL prefactor $\alpha$ correlating with the band-inversion transition. Beyond the diffusive limit, transport crosses into variable-range hopping, accompanied by a striking reversal of magnetoconductance from negative to positive. The observed positive low-field magnetoconductance, its pronounced anisotropy, and its temperature evolution point to an orbital origin of the response. These features are naturally captured by incorporating the incoherent hopping mechanism of Raikh \textit{et al.} together with wavefunction shrinkage, rather than by conventional quantum-correction frameworks. Our results provide a unified picture of how topology, spin-orbit coupling, and disorder collectively determine the full field-temperature magnetotransport landscape in this material class, establishing a clear experimental link between the topological phase transition and the onset of incoherent hopping-dominated conduction.
\end{abstract}

\maketitle

\section{\label{sec:level1}Introduction:}

Topological insulators (TIs) are a distinct class of quantum materials that host symmetry-protected surface states within a bulk energy gap~\cite{hasan2010colloquium,moore2010birth,qi2010quantum,ando2013topological}. These states are characterized by a nontrivial $Z_{2}$ topological invariant~\cite{kane2005z,fu2007topological,moore2007topological,fu2007topological.}, which distinguishes them from trivial band insulators in a manner analogous to the role of the Chern number in the quantum Hall effect \cite{haldane1988model,thouless1982quantized}. Unlike the quantum Hall state, which requires strong magnetic fields to emerge, the topological insulating state arises from strong spin-orbit coupling that induces a band inversion in certain materials~\cite{bernevig2006quantum,bernevig2006quantum.,fu2007topological}. The resulting surface states exhibit linear Dirac dispersion and spin-momentum locking \cite{hsieh2008topological,xia2009observation,pan2011electronic,zhang2009topological,hsieh2009tunable,hsieh2009observation}, rendering them robust against backscattering \cite{roushan2009topological} and, in the clean limit, resistant to strong localization effects such as Anderson localization \cite{groth2009theory,li2009topological}. Importantly, tuning parameters such as chemical composition, thickness, or spin-orbit strength can induce bulk bandgap closure and reopening, thereby driving a quantum phase transition between distinct topological classes, as demonstrated in HgTe quantum wells \cite{bernevig2006quantum.} and in $\mathrm{Bi}_{1-x}\mathrm{Sb}_{x}$
\cite{hsieh2008topological}. In addition to this topological transition, electronic transport in disordered TIs can transition to a localized regime, where conduction exhibits the variable-range hopping (VRH) mechanism \cite{bhattacharyya2017evidence,ren2010large,brahlek2012topological,liao2015observation}. This dual sensitivity to both topology and disorder not only enables the exploration of fundamental physics such as band inversion and localization transitions, but also propells TIs as a versatile platform for spin-polarized transport and low-dissipation electronic devices.

While a number of prior studies have examined compositional tuning and topological phase transitions in related systems such as $(\mathrm{Bi}_{1-x}\mathrm{In}_{x})_{2}\mathrm{Se}_{3}$, $(\mathrm{Bi}_{1-x}\mathrm{Sb}_{x})_{2}\mathrm{Te}_{3}$, and TlBi(S$_{1-x}$Se$_x$)$_2$, the emphasis has largely been on structural evolution, band inversion, and global transport signatures across the transition~\cite{brahlek2012topological,liu2013topological,ye2015origin,zhang2011band,xu2011topological,sato2011unexpected}. 
Most transport investigations rely on resistivity scaling, Hall measurements, or weak antilocalization (WAL) analyses to extract phase coherence and channel contributions~\cite{he2011impurity,checkelsky2009quantum,chen2010gate}. However, a detailed understanding of how magnetoconductance (MC) evolves from the diffusive regime into the strongly localized, variable-range hopping phase remains limited. Existing reports tend either to focus on the WAL regime near the topological side of the transition or to treat localization broadly in terms of bulk insulating behavior, without examining the full field and temperature dependence of magnetotransport~\cite{bhattacharyya2017evidence,ren2010large}. 
A systematic magnetoconductance analysis across different transport regimes is thus still lacking. Addressing this gap is crucial for disentangling the interplay between spin--orbit coupling, disorder, and incoherent hopping mechanisms, and for clarifying the connection between topology and localization phenomena in these materials.

In this work, we explore the evolution of transport properties in ${In_{x}}({Bi_{0.3}}{Sb_{0.7}})_{2-x}{Te_3}$ thin films by systematically tuning the composition parameter $x$. At $x = 0$, the system exhibits a topological insulating phase with bulk being insulating and the surface states exhibiting metallic character. With increasing $x$, it undergoes a quantum phase transition, evolving from a topological insulator to a trivial insulator. Substituting heavier Bi/Sb atoms with lighter In reduces the effective spin--orbit coupling, which drives a topological quantum phase transition through bulk bandgap closure and reopening with inverted ordering. This critical doping alters the topological invariant, transforming the system from a topologically nontrivial to a trivial phase. Beyond this transition, progressive doping enhances disorder and scattering, leading to a crossover from diffusive to localized transport, followed by the emergence of a VRH regime before the system evolves into a trivial band insulator. In the VRH regime, where electronic states are localized, we observe a positive MC that cannot be explained by the conventional quantum corrections associated with weak antilocalization (WAL) or weak localization (WL) \cite{hikami1980spin,he2011impurity}. The temperature dependence of the MC shows that the positive contribution observed at low magnetic fields can be attributed to interference effects between forward and time reversed hopping paths. However, at higher fields, the usual negative MC contribution arising from wavefunction shrinkage \cite{shklovskii2013electronic} becomes increasingly suppressed, indicating the presence of an additional mechanism. To account for this behavior in the strongly localized regime, we refer to the incoherent mechanism proposed by Raikh \cite{raikh1990incoherent,raikh1992mechanisms}, which provides a consistent framework for understanding these experimental trends. The temperature and magnetic field dependence of electron transport across these regimes provides insight into the interplay of topology, spin-orbit coupling, and disorder in these classes of materials.

\section{\label{sec:level1}Experimental Method}

Thin films of ${In_{x}}({Bi_{0.3}}{Sb_{0.7}})_{2-x}{Te_3}$ with a nominal thickness of 100 nm were synthesized on single-crystal $Al_2O_3$ substrates by pulsed laser deposition (PLD). Owing to its highly nonequilibrium nature, PLD allows the stoichiometry of the target material to be largely preserved in the deposited films. The targets were prepared from high-purity (99.999\%) Bi, Sb, Te, and In elements. A KrF excimer laser with a wavelength of 248 nm was employed at a repetition rate of 2 Hz. Prior to deposition, the chamber was evacuated to a base pressure of $6\times10^{-6}$ mbar, after which an argon atmosphere was introduced and maintained at a partial pressure of $4.5\times10^{-1}$ mbar. The substrate was heated to $240^\circ$C during growth, while the laser fluence was adjusted to approximately 1.2 J/cm$^2$ to ensure stable ablation. Following deposition, the films were subjected to an in-situ annealing process at the same substrate temperature for 20 minutes at base pressure. This post-deposition treatment improved the crystallinity and minimized surface roughness, thereby enhancing the overall structural quality of the thin films.  

\begin{figure}
\includegraphics[scale=0.35 ]{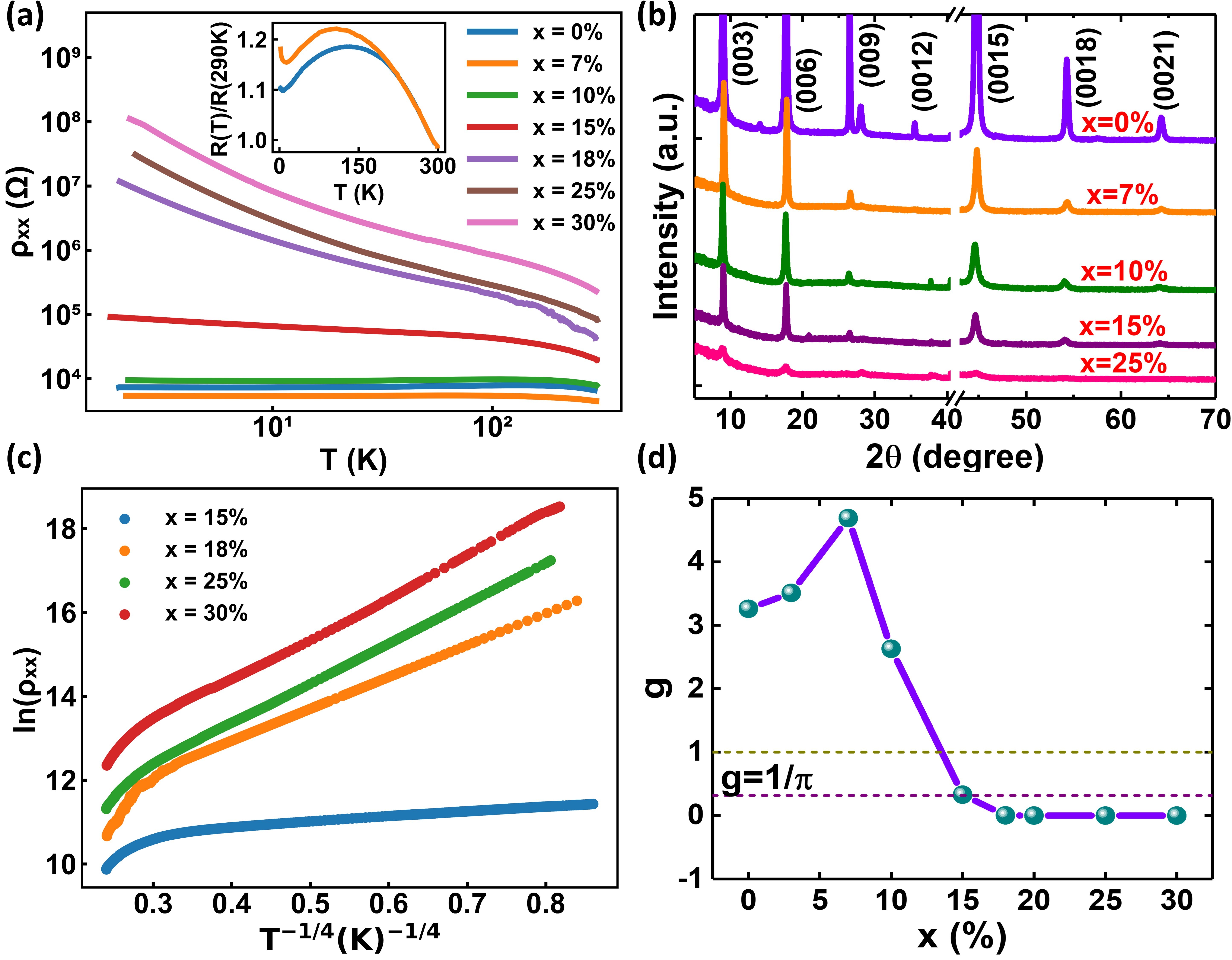}
\caption{\label{fig:1}(a) Temperature dependence of the resistance for ${\rm In}_{x}({\rm Bi}_{0.3}{\rm Sb}_{0.7})_{2-x}{\rm Te}_3$ thin films with varying In concentration $x$. The inset shows $R(T)/R(290\,{\rm K})$ for $x=0\%$ and $7\%$, both exhibiting metallic behavior at low temperatures. (b) X-ray diffraction patterns for $x = 0\%, 7\%, 10\%, 15\%, 25\%$, where the dominant $(003n)$ reflections confirm highly oriented $c$-axis growth. (c) Scaling of resistance with temperature, where the linear behavior indicates a 3D Mott VRH dependence for films with $x \ge 15\%$. (d) Dimensionless Drude conductivity illustrating the crossover from diffusive to localized transport. The threshold conductivity $g = 1/\pi$ marks the transition from negative to positive magnetoconductance as $x$ increases.}
\end{figure}

\section{\label{sec:level1}Results and Discussion:}

The resistance versus temperature R(T) curves Fig.~\ref{fig:1} reveal the evolution of the system from a topologically non-trivial to a topologically trivial insulating state as the In concentration (x) increases . Samples with x$<$ 15\% exhibit metallic behavior at low temperatures Fig.~\ref{fig:1}(a) inset, where decreasing resistance indicates the presence of robust, delocalized surface states that remain resilient against disorder and other interactions. This behavior reflects the enhanced contribution of surface-mediated carrier transport in the low-temperature regime, which is characteristic of a topological insulator \cite{gopal2017topological}. In contrast, for samples with x$\ge$ 15\%, the R(T) curves show an upward trend with decreasing temperature, signaling the onset of insulating behavior Fig.~\ref{fig:1}. For samples with x$<$ 15\%, a parallel resistor model was used to fit conductance versus temperature data (see the Supplemental Material for details). In this model, the total conductance (${\sigma}_{total}$) is expressed as the sum of the bulk conductance (${\sigma}_{bulk}$) and the surface conductance (${\sigma}_{surface}$), given by: 

\begin{equation} \label{eq1}
\begin{split}
{\sigma}_{total} &= {\sigma}_{bulk} + {\sigma}_{surface} \\
 & = \frac{1}{R_{0}exp^{\frac{\Delta}{T}}} + \frac{1}{A + BT}
\end{split}
\end{equation}

 where ${\sigma}_{bulk}$ = $\frac{1}{R_{0}exp^{\frac{\Delta}{T}}}$ and ${\sigma}_{surface}$ = $\frac{1}{A + BT}$. Here, $\Delta$ represents the activation energy of the impurity bands in bulk. The coefficient A corresponds to the temperature-independent disorder scattering, while the coefficient B accounts for the electron-phonon scattering \cite{xu2014observation,gopal2017topological}. At higher temperatures, conductance is primarily driven by thermally activated carriers originating from impurity states. As the temperature decreases, these bulk carriers get frozen out, resulting in surface states becoming dominant contributors to conductance in the thin films.

For the surface states of a topological insulator (TI), the $\pi$ Berry phase arising from spin-momentum locking leads to destructive interference in electron backscattering, resulting in weak antilocalization (WAL), a phenomenon opposite to weak localization (WL). Consequently, one expects the conductivity to increase with decreasing temperature, following a logarithmic (lnT) dependence \cite{garate2012weak}. However, electron-electron (e-e) interactions introduce a larger, opposing lnT correction that reduces the conductivity \cite{chiu2013weak}. The interplay between WAL and e-e interactions explains the observed temperature dependence of conductivity in the low-temperature regime (see the Supplemental Material for details).

In samples with x$\ge$ 15\%, the measured longitudinal resistance per square $\rho_{xx}$ significantly exceeds the quantum resistance $h/e^2$ ($\approx$ 25.8 K$\Omega$), indicating a high degree of disorder in the electronic system. The dimensionless Drude conductivity, $g = \sigma/(e^{2}/h) = k_{F} l$, provides a measure of the transport regime, where $\sigma$ is the electrical conductivity, $k_{F}$ the Fermi wave vector, and $l$ the mean free path. For g$>>$1,  the system resides in the diffusive metallic regime \cite{mott1979electronic}, while for g$<<$1, it enters the strongly localized regime. As g $\sim$1 the system approaches the quantum limit of diffusive transport, where the resistance becomes comparable to $h/e^2$. For compositions with x$<$ 15\%, the maximum of the measured longitudinal resistance per square $\rho_{xx,max}$ remains below $10~\mathrm{k}\Omega$, for which the value of dimensionless conductivity g$>>$1, indicating that the system resides in the diffusive transport regime. At x=15\%, $\rho_{xx,max}$ value is measured to be $\sim80~\mathrm{k}\Omega$ at 2K, corresponding to g $\sim$ 0.33 which signifies the crossover point between diffusive and localized transport. For compositions with x$>$ 15\%, $\rho_{xx,max}$ exceeds $1~\mathrm{M}\Omega$,placing the system well within the strongly localized regime, characterized by g$<<$1 Fig.~\ref{fig:1}(d). 

\begin{figure}
\includegraphics[scale=1]{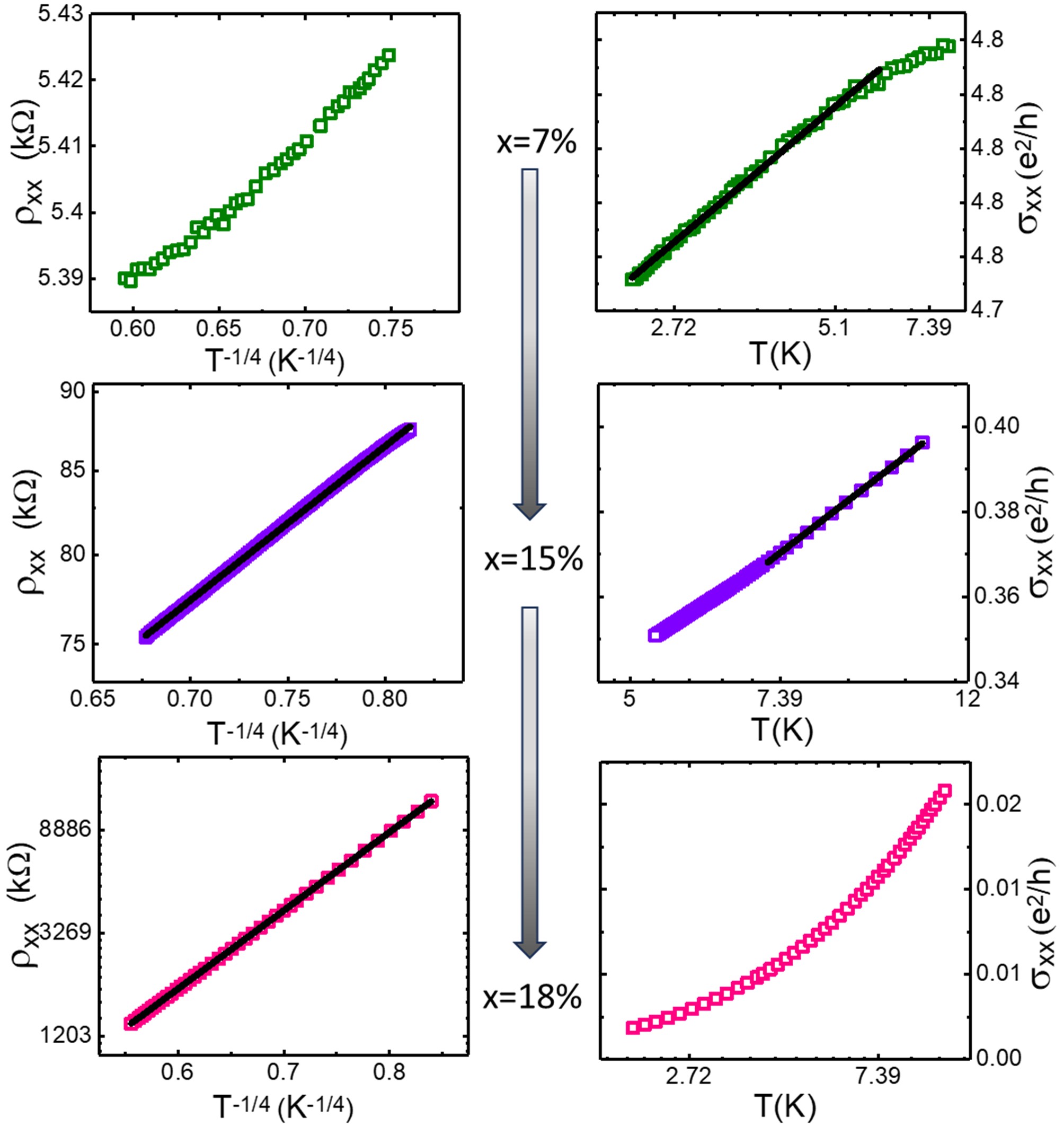}
\caption{\label{fig:2} The temperature dependence of $\rho_{xx}$ (in logarithmic scale) and $\sigma_{xx}$ (in linear scale) are plotted as function of $T^{-1/4}$ and T (in logarithmic scale) respectively. The difference in the x scale of the two sets of figures exhibits their respective temperature dependence. As disorder (doping concentration $x$) in the electronic system increases, the transport mechanism evolves from weak antilocalization to Mott variable-range hopping. Black solid lines are the liner fitting.}
\end{figure}

\begin{figure*}
\centering
\includegraphics[scale=1.05]{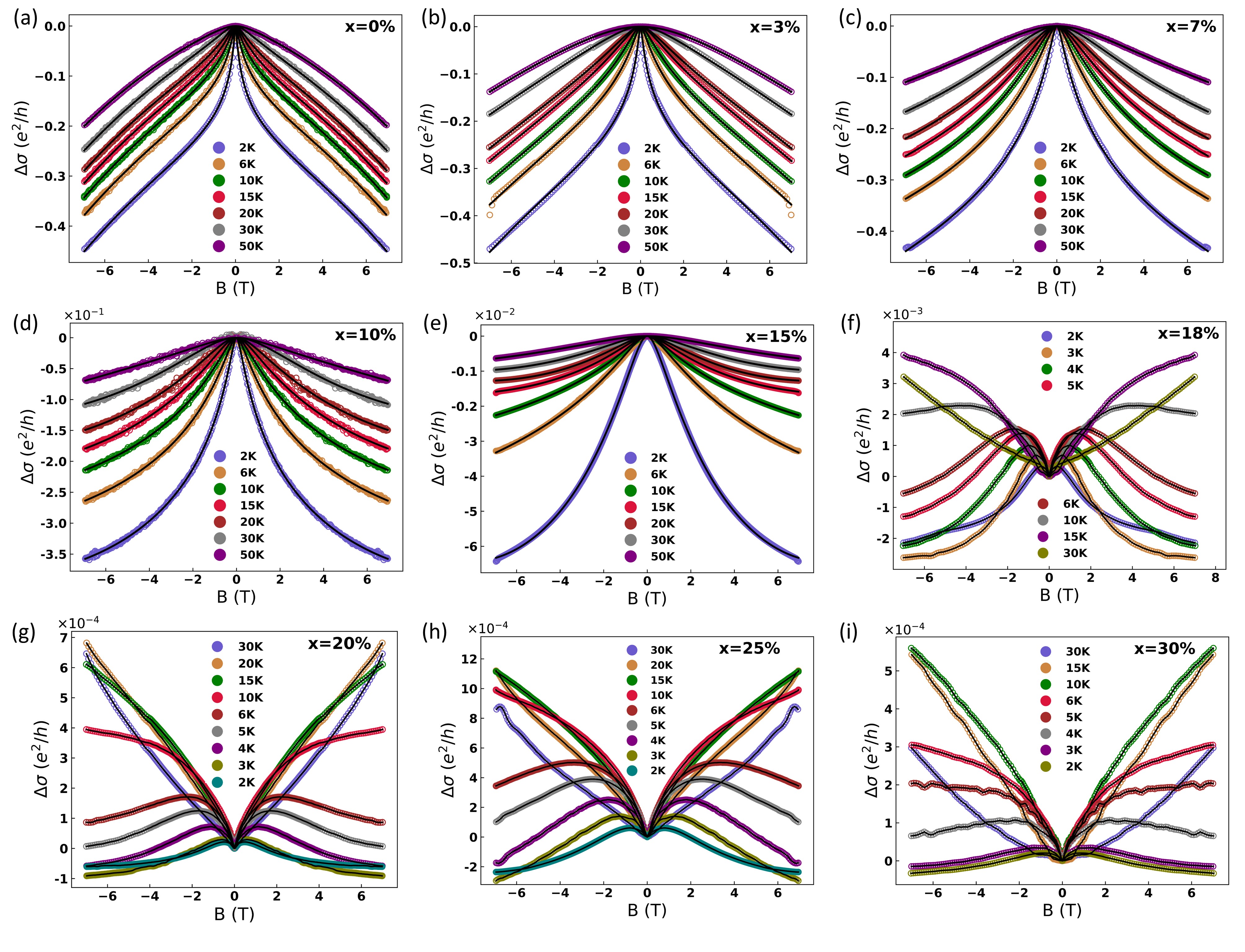}
\caption{\label{fig:3}Magnetoconductivity curves as a function of temperature for different doping percentages $x$. In Fig. (a–e), the solid black lines are fits using Eq. (3), combining the HLN and quadratic terms. In Fig. (f–i), the solid black lines are guides to the eye.}
\end{figure*}

The transition in the electronic transport regime with increasing disorder is clearly manifested in the temperature dependence of the conductivity as shown in Fig.~\ref{fig:2}. In the diffusive regime, where the dimensionless Drude conductivity is large (g$>>$1), the conductivity displays a logarithmic temperature dependence $\sigma(T)\propto ln(T)$ at very low temperatures, consistent with quantum corrections arising from weak localization and electron-electron interaction effects \cite{lu2014finite,altshuler1985electron,gopal2017topological}. In contrast, when the disorder becomes strong such that (g$<<$1) the system enters the strong localization regime, and transport is governed by the Mott variable range hopping (VRH) mechanism \cite{mott1979electronic}. In this regime, the temperature dependence of resistivity follows Mott’s variable-range hopping law,
\begin{equation}
    \rho(T) = \rho_{0} \exp\!\left[ \left( \frac{T_{0}}{T} \right)^{\tfrac{1}{d+1}} \right]
\end{equation}
where, $\rho(T)$ denotes the resistivity at temperature $T$, $\rho_{0}$ is the prefactor, $T_{0}$ is the characteristic Mott temperature related to the localization length and the density of states, and $d$ represents the dimensionality of the system (1, 2, or 3). In the intermediate regime ($g\sim 1$, x=15\%) corresponding to the crossover region between diffusive transport and strong localization, the temperature dependence of conductivity evolves from a logarithmic behavior  to an exponential form characteristic of Mott VRH as temperature decreases, as shown in Fig.~\ref{fig:2} . This transition can be understood in terms of the dominant dephasing mechanisms at play. At higher temperatures, dephasing is primarily governed by electron-phonon scattering. However, as the temperature decreases, the phonon contribution diminishes, and electron-electron (e–e) interactions become the dominant source of dephasing \cite{altshuler1985electron,lee2012gate,cha2012weak}. In this regime, the phase coherence length follows the relation $l_\phi\propto T^{-1/2}$, where $l_\phi=(D\tau_{\phi})^{-1/2}$,with D being the diffusion constant and $\tau_{\phi}$ the dephasing time. The localization length $\zeta$ denotes the characteristic decay length of the electronic wavefunction in a disordered potential, describing the spatial range over which quantum states remain coherent before becoming exponentially decaying. The relative magnitude of $l_\phi$ and the localization length $\zeta$ determines the nature of the electronic transport. When $l_\phi<<\zeta$, the system exhibits WAL, a quantum interference effect characteristic of the diffusive regime. In contrast, when $l_\phi>>\zeta$, electrons become fully phase coherent over distances larger than the localization length, and the system enters a strongly localized regime. As the temperature is lowered and $l_\phi$ increases beyond $\zeta$, the system undergoes a crossover from WAL behavior to strong localization, accompanied by a suppression of corrections to the quantum interference and the emergence of hopping-dominated conduction.

\begin{figure*}
\centering
\includegraphics[scale=0.5]{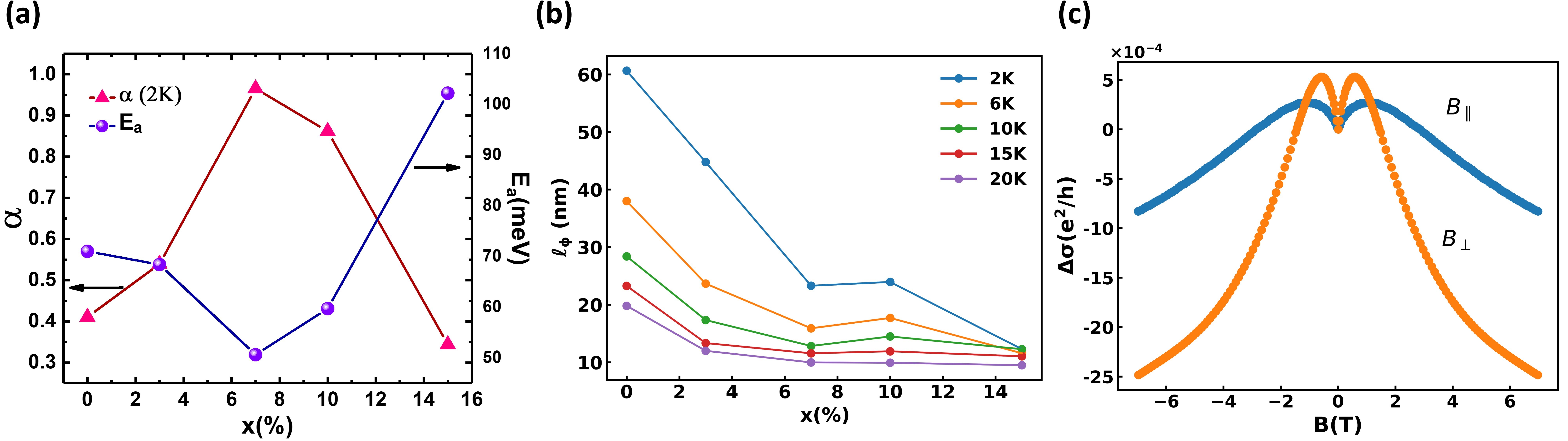}
\caption{\label{fig:4}(a) Hikami–Larkin–Nagaoka (HLN) fitting parameter $\alpha$ extracted at $T=2$~K as a function of doping concentration $x$, together with the activation energy obtained from Arrhenius fits to the temperature-dependent resistance in the diffusive transport regime. (b) Temperature dependence of the phase-coherence length $L_{\phi}$ for different doping concentrations $x$, determined from HLN analysis of the weak antilocalization magnetoconductance. 
(c) Magnetoconductance (MC) measured under in-plane and out-of-plane magnetic field configurations for $x = 18\%$, highlighting the pronounced anisotropic response.}
\end{figure*}

A crossover in the transport regime is typically reflected in a corresponding change in magnetoconductance (MC), defined as $\Delta\sigma=\sigma(B)-\sigma(0)$, as shown in Fig.~\ref{fig:2}. In low-disorder samples ($x<15\%$, with $g>>1$), the MC is negative, primarily due to the suppression of WAL by the applied magnetic field Fig.~\ref{fig:2}(a--e) . As the disorder increases ($x>15\%$, corresponding to $g<<1$), the system enters the strongly localized regime, and the MC exhibits a positive response in the low field limit Fig.~\ref{fig:2}(f--i) . For films in the diffusive regime, the observed negative MC can be quantitatively modeled by 

\begin{equation}
    \Delta\sigma(B) = -\alpha \frac{e^2}{2\pi^2\hbar} \left[ \psi\left(\frac{1}{2} + \frac{B_\phi}{B} \right) - \ln\left(\frac{B_\phi}{B}\right) \right] + \beta B^{2}
\end{equation}

where the first term is the Hikami–Larkin–Nagaoka (HLN) equation \cite{hikami1980spin}, which explains the negative MC resulting from the quantum correction due to WAL. The coefficient $\beta$ of the quadratic term reflects contributions from the elastic and spin-orbit scattering processes. In addition to these quantum effects, $\beta$ also includes a contribution from the classical cyclotronic MC \cite{assaf2013linear}. In HLN equation $\psi$ is the digamma function, $B_\phi = \frac{\hbar}{4 e l_\phi^2}$ is the dephasing field and $l_\phi$ is the phase coherence length. The coefficient $\alpha$ serves as an indicator of the effective number of two-dimensional (2D) conduction channels and reflects the underlying dimensionality of electronic transport in the system. In TI thin films, an ideal value of $\alpha$=1 is expected when the top and bottom surface states contribute independently, each providing a contribution of 1/2. However, experimental observations often reveal deviations from this idealized value, with $\alpha$ ranging from approximately 0.3 to 1.5 \cite{steinberg2011electrically,taskin2012manifestation,kim2011thickness,kim2013coherent,brahlek2014emergence,taskin2011berry,chen2010gate}. These variations are generally attributed to the presence of multiple parallel conduction channels, including contributions from surface states, bulk carriers, and possible interface or impurity-related pathways. HLN fitting has been performed for samples with compositions up to 15\%, and the extracted fitting parameters are presented in Fig.~\ref{fig:4}.

In systems with strong spin-orbit coupling, increasing disorder has previously been shown to induce a transition from negative MC, characteristic of WAL, to positive MC, indicative of strong localization. Hsu and Valles, in their experiments on ultrathin metallic films, demonstrated that this transition is governed by a well-defined threshold conductivity $\sigma_0=e^2/\pi\hbar $, (i.e $g=1/\pi$) \cite{hsu1995observation}. This critical value effectively distinguishes samples exhibiting negative MC behavior from those displaying positive MC Fig.~\ref{fig:1}(d). In our study, the low-field MC is observed to be negative for the \(x = 15\%\) sample and positive for the \(x = 18\%\) sample, indicating a sign reversal across this composition range. The transition occurs near the composition \(x = 15\%\), where the maximum resistivity reaches \(\rho_{xx,\mathrm{max}} \sim 80~\text{k}\Omega\) at 2~K, corresponding to a dimensionless conductivity \(g \sim 0.33\), close to the critical threshold value. This explains the observed change in MC from negative (for x$\le$15\%) to positive (for x$>$15\%), as the system crosses the threshold conductivity at low temperatures, indicative of a transition from WAL to strong localization.

From the magnetoconductivity fitting in the diffusive regime, we find that the parameter $\alpha$ increases with indium concentration $x$ up to about $7\%$, and then decreases as the system approaches the quantum limit of diffusive transport Fig.~\ref{fig:4}(a). A similar trend is observed in the activation energy obtained from Arrhenius fitting of the resistance–temperature data, where the activation energy decreases with increasing $x$ up to about $7\%$ and subsequently increases at higher concentrations. These correlations in $\alpha$ and the activation energy can be interpreted as signatures of a topological phase transition occurring near the critical concentration $x \approx 7\%$, where the band inversion takes place. This transition, marked by the closing and reopening of the bulk band gap and a corresponding change in the topological class of the system, was experimentally observed by Brahlek \textit{et al.}, who investigated transport and ARPES characteristics in thin films of $(\text{Bi}_{1-x}\text{In}_x)_2\text{Se}_3$ and showed that for indium concentrations of $3\%$--$7\%$, the system undergoes a band inversion accompanied by the reopening of the bulk gap~\cite{brahlek2012topological}. A similar trend has been reported by Liang Wu \textit{et al.}, who employed time-domain terahertz spectroscopy to probe the low-frequency conductance in $ (\mathrm{Bi}_{1-x}\mathrm{In}_x)_2\mathrm{Se}_3 $ \cite{wu2013sudden}. Their study revealed that beyond a certain substitution level, the transport lifetime collapses, signifying the breakdown of the topological phase. In addition, the emergence of a peak in the mid-infrared absorption with increasing indium concentration was shown to be consistent with a topological phase transition driven by the closing and reopening of the bulk band gap. It is worth noting that the critical concentration $x \approx 7\%$, associated with band inversion and a change in topological class, is distinct from the higher concentration threshold ($x \approx 15\%$), where the system crosses over from diffusive to strongly localized transport.

In addition to the influences of spin–orbit coupling and variations in the lattice constant, the bonding orbitals also change with $x$: Bi ($6p^3$) and Sb ($5p^3$) contribute only $p$ orbitals to bonding, whereas In ($5s^2 5p^1$) involves both $s$ and $p$ orbitals. This difference in orbital hybridization can significantly influence the topological phase transition.

\begin{figure*}
\centering
\includegraphics[scale=1]{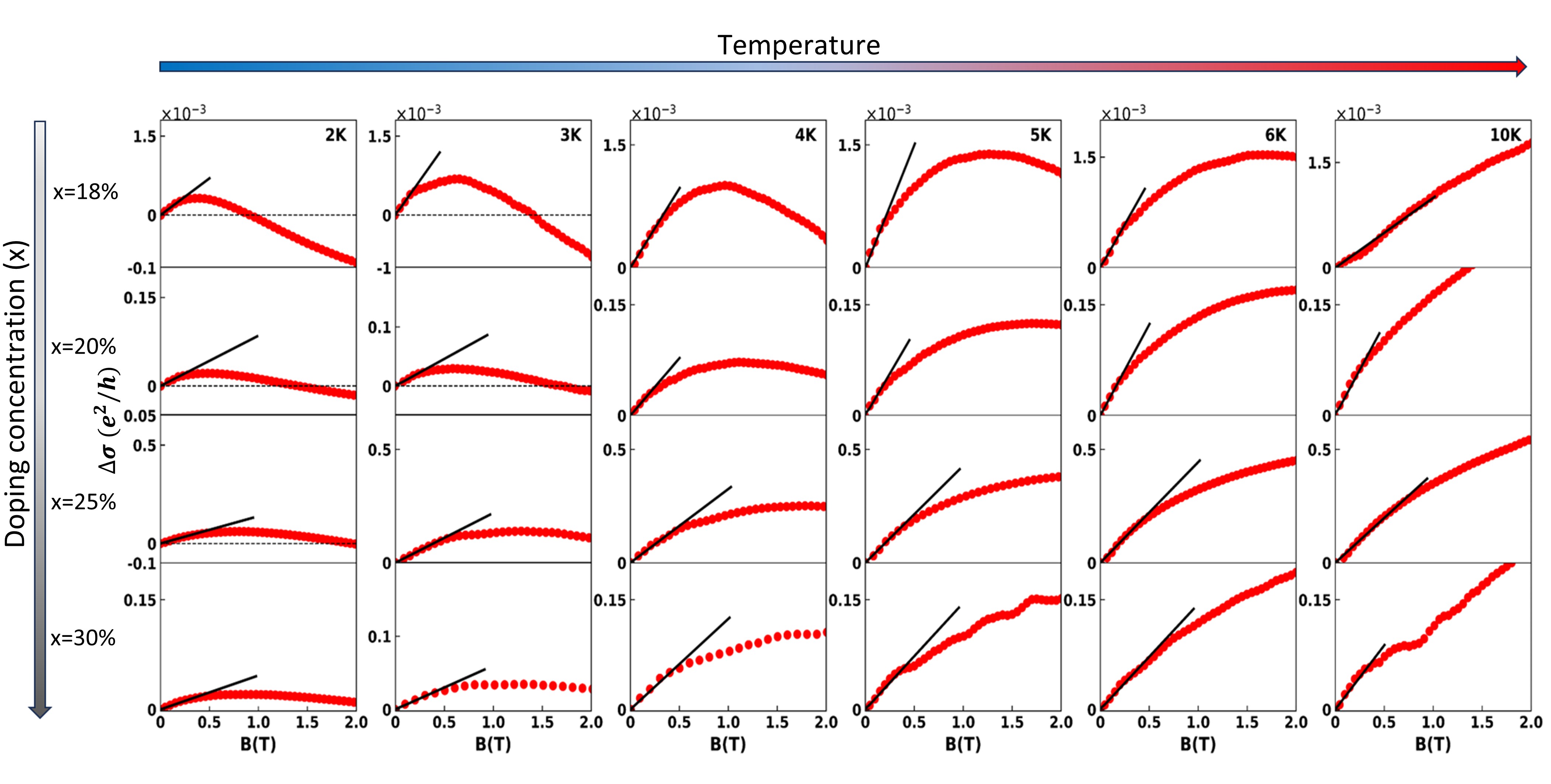}
\caption{\label{fig:5} Representative low field magnetoconductivity curves for different temperature and doping percentage $x$. Solid black line are the linear fit in the low field regime.}
\end{figure*}

In the localized regime ($x>15\%$), a positive MC is consistently observed at low magnetic fields across the temperature range where VRH conduction dominates Fig.~\ref{fig:5}. As the magnetic field increases, a crossover from positive to negative MC is observed. At very low temperatures, this crossover occurs at relatively small magnetic fields, whereas with increasing temperature the crossover field shifts to higher values for a given doping concentration $x$ Fig.~\ref{fig:5}. At sufficiently high temperatures within the VRH regime, the MC remains positive over the entire measured magnetic field range. Furthermore, at a fixed temperature, the crossover field is found to increase with increasing $x$. To understand the origin of the observed positive MC, angular-dependent measurements were performed Fig.~\ref{fig:4}(c). Orbital contributions to MC are inherently anisotropic, whereas spin-related mechanisms, such as Zeeman splitting \cite{fukuyama1979negative} and spin-flip scattering \cite{osaka1979theory}, are isotropic. The observed directional dependence of MC on the applied magnetic field orientation strongly suggests an orbital origin, effectively ruling out a spin-related mechanism.

In disordered systems where charge transport occurs via VRH, the application of a weak magnetic field can enhance conductivity, resulting in positive MC. This effect is attributed to a mechanism first proposed by Nguyen et al. \cite{nguen1985tunnel}, who demonstrated that, due to the long-range hops involved in VRH conduction, the scattering of tunneling electrons by intermediate impurities plays a significant role. This is different from the quantum corrections associated with WAL and weak localization (WL), where the applied magnetic field suppresses the destructive (in WAL) or constructive (in WL) interference of backscattered electron paths. This phenomenon is particularly relevant at low temperature part of the VRH regime, where the Mott hopping length significantly exceeds the average distance between impurities. Under such conditions, numerous impurity sites with energies far from the Fermi level are located near the path connecting two hopping sites. These impurities act as scattering centers for tunneling electrons. The scattering amplitudes contributed by these centers can be either positive or negative, depending on their energy level positions with respect to the fermi level. The overall tunneling amplitude is the sum of contributions from different scattering sequences, and due to the random signs of these partial amplitudes, destructive interference can occur between alternative paths connecting the initial and final states, leading to increased resistance. However, when a magnetic field is introduced, electrons gain an additional phase as they traverse different paths \cite{aharonov1959significance}. This phase disrupts the coherence of the interference pattern, thereby diminishing its effect. As a result, the likelihood of successful tunneling events increases, effectively lowering the resistance. It is important to note that this effect is significant only in weak magnetic fields, where the shrinkage of electronic wavefunctions can be neglected. In the hopping regime, Nguen \textit{et al.}~\cite{nguen1985tunnel} showed that quantum interference between different hopping trajectories separated by the optimal hopping distance $R_{\mathrm{hop}}$ modifies the tunneling probability at low magnetic fields. Since the total resistance of a hopping path is a product of many individual hopping events, the physically relevant quantity is the configurational average of $\ln \sigma$, which captures the influence of the rare, high-resistance links that dominate the overall transport. Their numerical evaluation of this logarithmic average leads to a linear negative magnetoresistance in the low-field limit. Sivan \textit{et al.}~\cite{sivan1988orbital} used a critical-path approach, where transport is controlled by the most resistive “bottleneck’’ hops in the percolation network. They showed that the magnetic field mainly affects these crucial hops, giving rise to a quadratic field dependence of the negative magnetoresistance within the corresponding low-field regime.

At high magnetic fields, the hopping conductivity in systems exhibiting VRH becomes strongly influenced by the magnetic field. This is primarily due to the effect arising from the magnetic-field induced shrinkage of the localized electronic wave function \cite{shklovskii2013electronic}. As the field strength increases, the spatial extent of the localized wave function becomes more confined. This shrinkage reduces the overlap between the wave functions of neighboring localized sites, which in turn decreases the probability of an electron successfully hopping from one site to another. The reduced overlap suppresses the conductivity, which causes the conductivity of the system to decrease with the magnetic field, which leads to a negative MC.

To explain the positive MC observed at high magnetic fields, we adopt the incoherent mechanism proposed by Raikh \cite{raikh1990incoherent,raikh1992mechanisms}. This mechanism attributes the effect to magnetic-field-induced changes in the energies of localized states. A magnetic field compresses the wave functions of localized impurity states. This reduction in spatial extent decreases their overlap with neighboring impurity sites, weakens the associated level repulsion, and consequently shifts the impurity energy levels. Since the energy of each conducting site is influenced by its interactions, mediated through wave function overlap, with the surrounding non-conducting impurities that form a background of localized states, any field-induced shifts in these energies consequently modify the hopping rates between conducting sites. As a result, both the activation energy that governs the hopping probability and the tunneling amplitude acquire a magnetic field dependence, thereby altering the overall electrical transport. Raikh derived a correction term for the activation energy by incorporating the overlap integral $V_{ij}$ of the impurity wave functions at sites $i$ and $j$. The resulting correction, $\delta \Delta _{ij}$, is positive, leading to a suppression of hopping relative to the case where the overlap integral is neglected. Thus, the overlap between the wave functions of the impurities and their nearest neighbors decreases the likelihood of hopping. However, in the presence of a magnetic field, the absolute magnitude of $V_{ij}$ decreases as a result of the shrinkage of the wave function. Consequently, the correction $\delta \Delta _{ij}$ is reduced, effectively lowering the activation energy and increasing the probability of hopping, resulting in increased conductivity. Thus, this mechanism provides a route to positive MC at high magnetic fields. Importantly, it competes with the conventional negative MC mechanism, also arising from wave function shrinkage, thereby indicating the coexistence and interplay of the two mechanisms.

\begin{figure}
\includegraphics[scale=0.28]{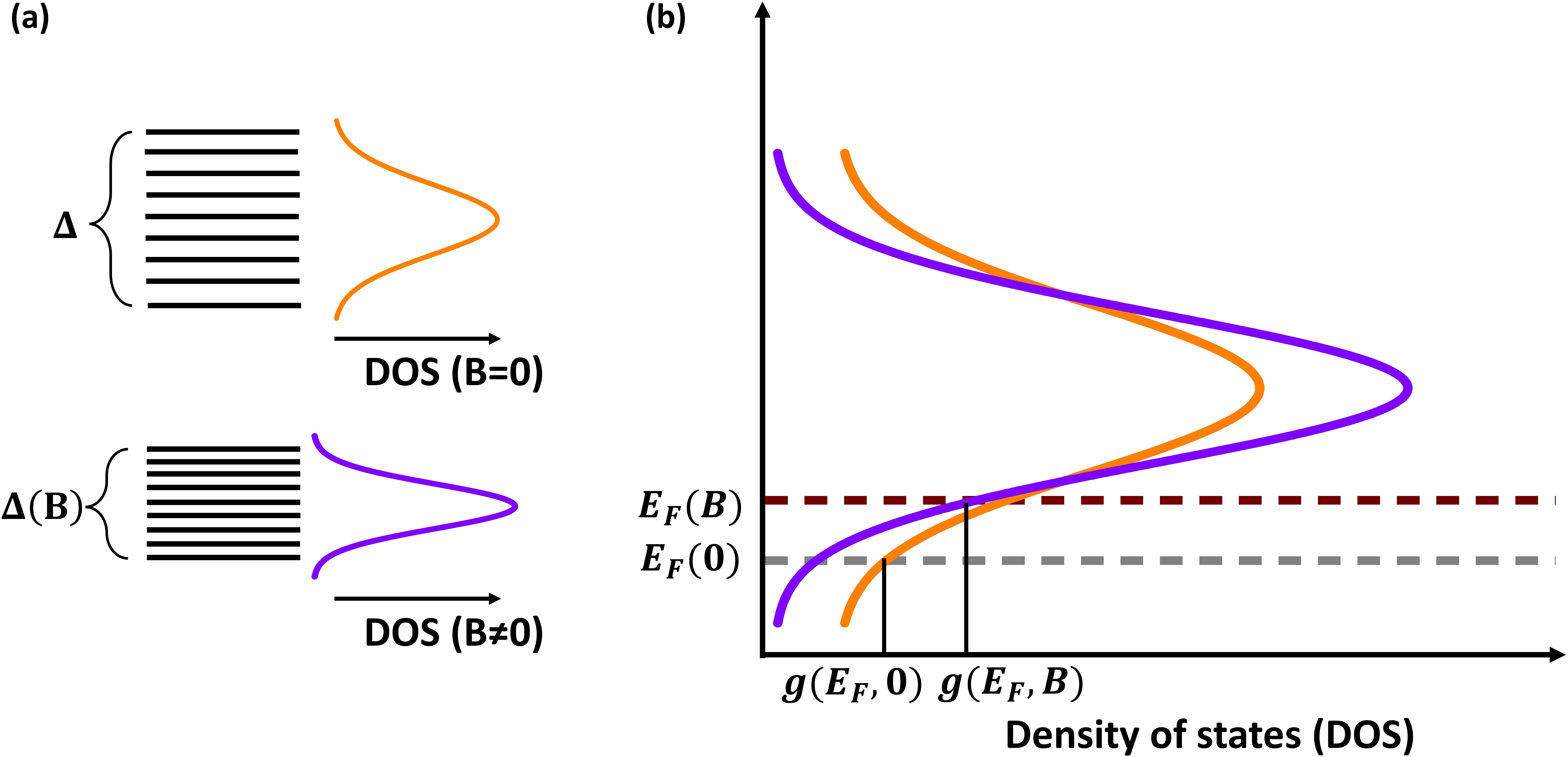}
\caption{\label{fig:6} (a) Schematic illustration of the impurity-band width $\Delta$. An applied magnetic field reduces the overlap between neighboring impurity wave functions, suppressing impurity level broadening and thereby narrowing the impurity band. The corresponding modification of the density of states (DOS) is shown. (b) Magnetic-field–induced band narrowing drives a shift of the Fermi level toward the band center to conserve the total carrier density, leading to an enhanced density of states at the Fermi energy. }
\end{figure}

The dependence of MC on temperature and doping in the localized regime is illustrated in the Fig.~\ref{fig:5}. At low magnetic fields, the observed positive MC can be attributed to the quantum interference mechanism proposed by Nguyen et al., which yields a linear dependence on $B$ as shown in Fig.~\ref{fig:5}. At a fixed temperature, the crossover field is found to increase systematically with increasing doping concentration ($x$). Enhanced doping introduces additional nearby impurities that, while not directly involved in the hopping process, influence the activation behavior as proposed by Raikh. Under an applied magnetic field, the overlap between neighboring impurity wave functions is reduced, thereby suppressing the broadening of impurity energy levels and leading to a narrowing of the impurity band. This will result in a change in the density of states of the impurity band \cite{raikh1992mechanisms}. As the field strength increases, the Fermi level shifts toward the band center in order to keep the total number of carriers conserved as shown in Fig.~\ref{fig:6}. If the density of states in the tail decreases sufficiently rapidly, this shift enhances the density of states at the Fermi level, thereby lowering the resistance of the system. Such an enhancement in the density of states can be inferred from the magnetic-field dependence of the Mott characteristic temperature $T_{0}$ (see the Supplemental Material for details). Consequently, with increasing doping concentration, the negative MC mechanism is activated only at higher magnetic fields, since the denser impurity landscape suppresses wave-function shrinkage and localization at lower fields. As a result, the crossover field shifts to higher values. For a given doping concentration, the crossover field increases with temperature, and MC becomes positive at higher temperatures (within the temperature range where VRH remains the dominant conduction mechanism). Band narrowing under a magnetic field shifts the Fermi level within the impurity band, thereby increasing the density of states at the Fermi energy, which in turn increases the conductivity. At low temperatures, hopping mainly involves states very close to the Fermi energy. As the temperature increases, thermal smearing broadens the energy window of accessible states around the Fermi level (occupation around the Fermi level). This allows electrons to access states closer to or even above the mobility edge, where localization is weaker or extended states exist. With increasing temperature, the positive MC mechanism proposed by Raikh \cite{raikh1990incoherent,raikh1992mechanisms} dominates over the negative MC contribution, owing to the enhanced density of states near the Fermi energy and the thermal broadening. Hence, the observed MC behavior reflects the magnetic field driven restructuring of the impurity band, which modulates both the available density of states and the character of hopping pathways.

\section{\label{sec:level1}Conclusion}

Our comprehensive study of ${In_{x}}({Bi_{0.3}}{Sb_{0.7}})_{2-x}{Te_3}$ thin films reveals a complex interplay between disorder, band topology, and quantum transport. From resistance and magnetoconductivity analyses, two characteristic doping levels emerge as key markers of the transport evolution. Around $x \approx 7\%$, the non-monotonic variation of the HLN parameter $\alpha$ together with the activation energy trend indicate a topological phase transition, associated with the closure and reopening of the bulk band gap and the resulting change in the system’s topological class. At higher doping levels ($x \approx 15\%$), the system transitions from diffusive transport to strong localization, marked by resistivity exceeding the quantum resistance threshold. For $x < 15\%$, MC is negative and is largely governed by bulk scattering process and surface states that display weak antilocalization, whereas For $x > 15\%$, the MC is positive at low fields but turns negative at higher fields, while at elevated temperatures it remains positive across the entire field range. This crossover can be understood in terms of the interplay between the path-interference mechanism of Nguyen \cite{nguen1985tunnel}, relevant at weak fields, and the incoherent process proposed by Raikh \cite{raikh1990incoherent}\cite{raikh1992mechanisms} along with wave function shrinkage, which dominates at stronger fields. Taken together, these findings build a coherent framework for understanding how band inversion, increasing disorder, and quantum interference effects collectively dictate transport in indium doped topological insulators, thereby deepening our understanding of disorder-driven topological transitions.

\bibliography{apssamp}% Produces the bibliography via BibTeX.

\end{document}